\documentclass[12pt,a4paper]{article}
\usepackage{amsmath,amssymb,bm,ascmac,bbm,mathtools}
\usepackage[dvipdfmx]{graphicx}
\graphicspath{./images/}
\usepackage{cite}		
\usepackage{xcolor}
\usepackage{here}
\usepackage{authblk}
\usepackage[hang,small,bf]{caption}
\usepackage[subrefformat=parens]{subcaption}
\usepackage{dcolumn}

\newcommand{\tr}{\text{tr}}

\newcommand{\mcal}{\mathcal}
\newcommand{\mbb}{\mathbb}

\newcommand{\End}{\text{End}}

\newcommand{\ep}{\epsilon}

\begin{document}
\title{Emergent SUSY in two dimensions}
\author{Ken Kikuchi}
\affil{Yau Mathematical Sciences Center,
Tsinghua University}
\date{}
\maketitle

\begin{abstract}
We propose a renormalization group flow with emergent supersymmetry in two dimensions from a non-Lagrangian theory. The ultraviolet theory does not have supersymmetry while the infrared theory does. We constrain the flow both analytically and numerically (truncated conformal space approach). Analytic constraints include a new spin constraint.
\end{abstract}

\makeatletter
\renewcommand{\theequation}
{\arabic{section}.\arabic{equation}}
\@addtoreset{equation}{section}
\makeatother

\section{Introduction}
Although supersymmetry (SUSY) has been playing important roles in mathematics, theoretical physics, and phenomenology, the symmetry has not been observed yet. Here, let us recall why we succeeded to discover Higgs(-like) particle. In that case, a first principle \textit{required} the presence of the particle. Namely, Higgs particle is needed to respect unitarity. This lesson raises a natural question: Can we find a situation where an existence of SUSY is forced by first principles? We propose one situation in two dimensions. The unitarity and symmetries \textit{require} an existence of SUSY.

More precisely, we propose a renormalization group (RG) flow with emergent SUSY; the ultraviolet (UV) theory does not have SUSY, while the infrared (IR) theory does. Concretely, we start from the fermionic $m=5$ minimal model \cite{P88,FGP90}, which was rediscovered recently \cite{RW20,HNT,K20} through fermionization \cite{KTT19,JSW19}. The UV theory is a strongly-coupled non-Lagrangian theory without SUSY. We deform the theory with a specific relevant operator. It triggers an RG flow. Along the flow, generalized symmetries are preserved, and their constraints together with unitarity claim the IR theory is the fermionic $m=4$ minimal model with $\mcal N=1$ SUSY \cite{FQS84}.\footnote{Previous proposals of emergent SUSY include the flow from the lattice model of interacting Majorana fermions \cite{RZFA15}, and the flow from the two-dimensional quantum chromodynamics to the ${\cal N}=(2,2)$ Kazama-Suzuki models \cite{DGY21}. There is another proposal of emergent SUSY on the edge \cite{GV12,DGG21,BL21,CHW22}. See also \cite{T14,GKW18,BB18,GY18,MS16,Zafrir:2020epd} for related phenomena of SUSY enhancement in various dimensions.} In order to solve the RG flow, we employ not only analytic constraints from the generalized symmetries but also a numerical method, the truncated conformal space approach (TCSA). TCSA tells us the ground state degeneracies and the IR central charge that strongly support our flow.

\section{Constraining RG flows}
In a modern language \cite{GKSW}, symmetries are generated by topological operators supported on defects with certain codimension. Their properties are constrained by various consistency conditions. For instance, in two dimensions, the action of a (zero-form) symmetry on local operators is implemented by bringing a topological defect line (TDL) across the local operators. Their actions are constrained by the Cardy condition \cite{C86,PZ00}.\footnote{We focus on zero-form (possibly non-invertible) symmetry (also called category symmetry) in this paper.} The actions on the identity operator is called the quantum dimension. The fusions of the TDLs, corresponding to the composition of the symmetry actions (together with the direct sum) form fusion rings. The fusion is imposed to obey the pentagon axiom \cite{MS88,MS89,FRS02,FRS03,FRS04a,FRS04b,BT17,CLSWY}. The solution to the axiom is called $F$-symbols. In rational conformal field theories (RCFT), TDLs usually admit additional structure, braiding. It is imposed to obey the hexagon axiom \cite{MS88,MS89,FRS02,FRS03,FRS04a,FRS04b}. These constraints are so strong that the data of the TDLs --- the quantum dimension, fusion ring, $F$-symbols, and braidings --- can only take a discrete set of values.\footnote{This is known as the Ocneanu rigidity in category theory \cite{ENO02}.}

Consider an RG flow triggered by a relevant operator. It was realized that if a TDL commutes with the relevant operator, the TDL survives all along the RG flow. It was argued \cite{CLSWY} that the discrete data associated to such TDLs cannot be deformed continuously under the flow. A simple consequence following from the invariance of the quantum dimension is that if a TDL with a non-integer quantum dimension is preserved under the RG flow, then the vacuum cannot be gapped without degeneracy. The invariance of quantum dimensions is an immediate consequence of the invariance of (topological) $S$-matrices. The invariance also succeeded to explain symmetry enhancements in RCFTs \cite{K21}.

We give new analytic constraint on RG flows. Let us put a theory on a cylinder. A time slice is given by a circle. We have Hilbert space $\mcal H$ on it. Now, pick a TDL $\mcal L$. We can insert it along the time axis. The insertion modifies the original Hilbert space $\mcal H$ to a defect Hilbert space $\mcal H_{\mcal L}$. The states (or operators) in the defect Hilbert space have specific spins derived from modular transformations and the $F$-symbols \cite{CLSWY}. We denote the set of spins associated to $\mcal H_{\mcal L}$ as $S_{\mcal L}$ and call it the spin content. Suppose we deform a UV theory with spacetime scalar. RG flows triggered by such operators preserve rotation symmetry. Thus, its quantum numbers (i.e., spins) are conserved. Therefore, we claim spin contents $S_{\mcal L}$ associated to a surviving TDL $\mcal L$ should be invariant. Taking into account the possible decoupling of heavy operators, spin content $S_{\mcal L}^\text{IR}$ in the IR must be a subset of that in UV $S_{\mcal L}^\text{UV}$:

\begin{equation}
    S_{\cal L}^\text{IR}\subset S_{\cal L}^\text{UV}.\label{spinmatch}
\end{equation}
We employ this constraint\footnote{Note the difference between the spin selection rules \cite{CLSWY} and our spin constraints. Spin contents in all theories with the same fusion category should obey the same spin selection rules. In order to obtain the selection rules, one has to solve cumbersome pentagon identities. In this sense, the spin selection rules are theory-independent, though they are difficult to derive. In other words, one cannot get spin selection rules just from spin contents of some concrete theories. On the other hand, our spin constraints do not require $F$-symbols and hence avoid solving the cumbersome pentagon identities. In this sense, our spin constraint is theory-dependent, while they are easy to obtain. Although our spin constraint cannot tell about the other theories sharing the same fusion category, what one needs to know in constraining RG flows are the spin contents of the UV theory and those of the candidate IR theories. Furthermore, we made the implicit assumption in \cite{CLSWY} explicit; when one breaks rotation symmetry with spinning operators, there is no reason for the spin quantum numbers to be preserved.} below to exclude some putative IR theories.

\section{Application: an RG flow with emergent SUSY}
As an application of the constraints on RG flows, we consider a particular RG flow from the fermionic $m=5$ minimal model. The flow is interesting because one IR phase is expected to be the fermionic $m=4$ minimal model, which is an $\mcal N=1$ SCFT \cite{FQS84}. 

\subsection{Analytic constraints}
Now, let us pick the fermionic $m=5$ minimal model as the UV theory. The theory in the antiperiodic (Neveu-Schwarz) sector has ten primary operators 
\begin{equation}
1\,,\quad \epsilon\bar\epsilon\,,\quad \epsilon'\bar\epsilon'\,,\quad \epsilon''\bar\epsilon''\,,\quad \epsilon'''\bar\epsilon'''\,,\quad \epsilon''''\bar\epsilon''''\,,\quad \sigma\bar\sigma''\,,\quad \sigma''\bar\sigma\,, \quad \sigma'\bar\sigma'''\,,\quad \sigma'''\bar\sigma'\,,
\end{equation}
with conformal weights $(h,\bar h)=(0,0)$, $(\frac1{15},\frac1{15})$, $(\frac25,\frac25)$, $(\frac23,\frac23)$, $(\frac75,\frac75)$, $(3,3)$, $(\frac1{40},\frac{21}{40})$, $(\frac{21}{40},\frac1{40})$, $(\frac18,\frac{13}8)$, $(\frac{13}8,\frac18)$, 
respectively. The relevant scalar primaries are $\ep\bar\ep$, $\ep'\bar\ep'$ and $\ep''\bar\ep''$, that could be our candidates for triggering the RG flow.

By solving the Cardy condition, we find ten TDLs in the fermionic $m=5$ minimal model (see supplemental material \ref{TDLsinm=345}). The fusion ring is generated by four TDLs $N$, $M$, $(-1)^F$, $W$ with the relations
\begin{equation}\label{eqn:fusionrulem=5}
\begin{split}
    N^2=I+M\,,\quad M^2=I+M+(-1)^F\,,&\quad \left((-1)^F\right)^2=I\,,\quad (-1)^F M=M\,,
    \\
    NM=N+(-1)^FN\,,&\quad W^2=I+W\,.
\end{split}
\end{equation}
Their actions on the primary operators are given in Table \ref{TDLactionm=5}.
\begin{table}[H]
\begin{center}
\begin{tabular}{c|c|c|c|c|c|c|c|c|c|c}
	&$1$&$\epsilon\bar\epsilon$&$\epsilon'\bar\epsilon'$&$\epsilon''\bar\epsilon''$&$\epsilon'''\bar\epsilon'''$&$\epsilon''''\bar\epsilon''''$&$\sigma\bar\sigma''$&$\sigma''\bar\sigma$&$\sigma'\bar\sigma'''$&$\sigma'''\bar\sigma'$
	\\\hline
	$\widehat N$&$\sqrt3$&$0$&$-\sqrt3$&$0$&$\sqrt3$&$-\sqrt3$&$-1$&$1$&$1$&$-1$
	\\
	$\widehat M$&$2$&$-1$&$2$&$-1$&$2$&$2$&$0$&$0$&$0$&$0$
	\\
	$\widehat{(-1)^F}$&$1$&$1$&$1$&$1$&$1$&$1$&$-1$&$-1$&$-1$&$-1$
	\\
	$\widehat W$&$\zeta$&$-\zeta^{-1}$&$-\zeta^{-1}$&$\zeta$&$-\zeta^{-1}$&$\zeta$&$-\zeta^{-1}$&$-\zeta^{-1}$&$\zeta$&$\zeta$
\end{tabular}
\end{center}
\caption{The action of the primitive TDLs in the fermionic $m=5$ minimal model on the primary operators, where $\zeta:=\frac{1+\sqrt5}2$ is the golden ratio. The TDL $(-1)^F$ acts as the fermion parity, i.e. $+1$ on the bosonic operators and $-1$ on the fermionic operators.}\label{TDLactionm=5}
\end{table}

We will focus on the RG flow triggered by the operator $\ep''\bar\ep''$ and study the constraints coming from the surviving TDLs $\{I,\,(-1)^F,\,W,\,(-1)^FW\}$.\footnote{For the other two RG flows, see supplemental material \ref{theother2}.} The other relevant operators $\ep\bar\ep$ and $\ep'\bar\ep'$ cannot be generated along the RG flow because they are prohibited by the non-invertible symmetry $W$.

First, by the constraint of the quantum dimension discussed in the previous section, the IR theory cannot be trivially gapped, since the line $W$ has non-integer quantum dimension $\zeta\notin\mbb N$.
The possible IR phases are either a CFT or a topological quantum field theory (TQFT) with degenerated vacua.\footnote{Logically speaking, we cannot rule out a tensor product of a TQFT and a CFT where all the surviving TDLs flow to those in the TQFT factor, and the TDLs in the CFT factor are all emergent. However, this scenario requires many emergent lines without any reason (so far); thus, is unnatural \cite{K21}.} Furthermore, since the possible eigenvalues of $W$ are $\frac{1\pm\sqrt5}2$, the ground state degeneracy (GSD) of the TQFT should be even \cite{CLSWY}. Hence, the possible IR phases can be summarized as follows depending on the GSD:
\begin{equation}
\text{IR theory}=\begin{cases}
\text{CFT}&(\text{GSD}=1),\\
\text{TQFT}&(\text{GSD}\in2\mbb N).
\end{cases}\label{IRphases}
\end{equation}
In order to find the GSD, we employ a numerical method, the truncated conformal space approach (TCSA) \cite{YZ89}, in the next subsection. We will see $\text{GSD}=1$ for one sign of the relevant coupling, and $\text{GSD}=2$ for the opposite sign.

Let us first consider the CFT phase. From the $c$-theorem, there are only four candidates: bosonic or fermionic minimal models with $m=4,3$. They have central charges $c=\frac7{10}$ and $c=\frac12$, respectively. Note that the $m=3$ minimal models do not have a TDL with quantum dimension $\zeta$, so the IR theory cannot be the bosonic or fermionic $m=3$ minimal model. Therefore, we are left with fermionic or bosonic $m=4$ minimal models. 

Let us assume that the IR CFT is the bosonic $m=4$ minimal model.\footnote{The possibility of flows to bosonic minimal model was raised by Jin Chen.} Since the fusion rules of the surviving TDLs are preserved along the RG flow, the fermion parity $(-1)^F$ should flow to a TDL that generates a $\mbb Z_2$ symmetry. One may naively match it with the $\mbb Z_2$ line in the bosonic $m=4$ minimal model. However, this possibility can be ruled out by our new spin constraint \eqref{spinmatch}. The defect Hilbert space of the fermion parity line $(-1)^F$ has the spin content
\begin{equation}\label{eqn:fermion_parity_spin_content}
    s\in\{0,\pm1,\pm3\}\,,
\end{equation}
while the defect Hilbert space of the $\mbb Z_2$ line in the bosonic $m=4$ minimal model has the spin content
\begin{equation}
    s\in\{0,\pm\frac12,\pm\frac32\}\,,
\end{equation}
which is $not$ the subset of the former \eqref{eqn:fermion_parity_spin_content}. 
The remaining possibility is that $(-1)^F$ flows to a TDL that acts trivially on all the local operators in the bosonic $m=4$ minimal model. In this scenario, the $\mbb Z_2$ symmetry in the bosonic $m=4$ minimal model must emerge, which we found unnatural \cite{K21}. In the next section, we will provide numerical evidences for the existence of IR fermionic states, which would rule out the flow to IR bosonic phases.

Finally, let us examine the only  CFT scenario, the fermionic $m=4$ minimal model. By solving the (modified) Cardy condition, we find eight TDLs in the fermionic $m=4$ minimal model (see supplemental material \ref{TDLsinm=345}). The fusion ring is generated by the TDLs $(-1)^F$, $R$, $W$ with the relations
\begin{equation}
    \left((-1)^F\right)^2=I\,,\quad R^2=I\,,\quad W^2=I+W\,.
\end{equation}
By matching the UV and the IR fusion rules, we must have
\begin{equation}
    W_{\rm UV}=W_{\rm IR}\,.
\end{equation}
Next, let us match the UV TDL $(-1)^F$. There are three nontrivial $\mbb Z_2$ lines in IR: $(-1)^F$,  $R$, $(-1)^FR$. Can we figure out to which TDL $(-1)^F$ flows? Yes, we can. The defect Hilbert space $\mcal H_{(-1)^F}$ of the UV $(-1)^F$ contains only spin zero states, while those of the IR TDLs have the spin contents $s\in\{0\}$ for $(-1)^F$, $s\in\{-\frac{1}{16},\,\frac{7}{16}\}$ for $R$ and $s\in\{\frac{1}{16},\,-\frac{7}{16}\}$ for $(-1)^FR$. Our spin constraint \eqref{spinmatch} implies that the UV $(-1)^F$ can only flow to the IR $(-1)^F$.\footnote{The flows from the UV $(-1)^F$ to the IR $R$ or $(-1)^F R$ are also ruled out by matching the $F$-symbols.} This completes the matching of surviving TDLs because the fourth line should be the fusion of $(-1)^F$ and $W$.

In short, surviving lines are identified as
\begin{equation}
    \begin{array}{ccccc}
    \text{UV}:&I&(-1)^F& W&(-1)^FW\\
    &\downarrow&\downarrow&\downarrow&\downarrow\\
    \text{IR}:&I & (-1)^F&W&(-1)^FW
    \end{array}.\label{matchTDLs5to4}
\end{equation}
Note that the IR theory is an RCFT described by a modular tensor category (MTC). However, the surviving lines do $not$ form an MTC, but only a braided fusion category (BFC) because the $(-1)^F$ line is transparent.\footnote{The monodromy charge matrix is given by \cite{fMTC,KK}
\begin{equation}
    M=\begin{pmatrix}1&1&1&1\\1&1&1&1\\1&1&-\frac1{\zeta^2}&-\frac1{\zeta^2}\\1&1&-\frac1{\zeta^2}&-\frac1{\zeta^2}\end{pmatrix}\label{braiding}
\end{equation}
in the basis $\{I,(-1)^F,W,(-1)^FW\}$.} Hence, extra TDLs should emerge in the IR \cite{K21,KK}. In particular, the emergent TDLs $R$ and $(-1)^FR$ generate the ${\mathbb Z}_2\times {\mathbb Z}_2$ $R$-symmetry of the emergent $\mcal N=1$ SUSY. An important fact to notice is that the TDLs $R$ and $(-1)^FR$ are q-type \cite{ALW17,Zhou:2021ulc}, which satisfy the modified Cardy condition discussed in supplemental material \ref{TDLsinm=345}.

\subsection{Numerical study}
In the truncated conformal space approach (TCSA), one puts a CFT on a cylinder with circumference $R$. The theory is quantized on a circle with a Hamiltonian given by that of the UV CFT. One then deforms the Hamiltonian by adding terms given by integrating the relevant primary operators. The full Hamiltonian after the deformation is\footnote{We fix the sign of couplings in the Lagrangian formalism.}
\begin{equation}
    H_\text{full}=H_\text{UV CFT}-\sum_i\lambda_i\int_{\mbb S^1}\phi_i.\label{fullH}
\end{equation}
The spectrum of the deformed theory is obtained by diagonalizing the full Hamiltonian. Originally, this task is intractable due to the infinitely many states. Yurov and Zamolodchikov \cite{YZ89} suggested to truncate the conformal Hilbert space at some level. The truncation makes the dimension finite, and one can compute the spectrum. This is the reason for the name.

This problem can be solved on a computer. As far as we know, there are three open codes \cite{STRIP,TruSpace,HHT22} running on Mathematica, C++, and Matlab, respectively. Our code is based on STRIP \cite{STRIP}. We modify the code in two points: make it work also for nondiagonal theories, and made two improvements. The two improvements are 1) the subtraction of divergences as the cutoff in the TCSA is sent to infinity, and 2) the renormalization of coupling constant(s). For details of the improvements we use, see \cite{GW11,LT14}. 

\begin{figure}[H]
\begin{tabular}{cc}\hspace{-50pt}
    \begin{minipage}[t]{0.6\hsize}
    \centering
    \includegraphics[width=0.8\textwidth]{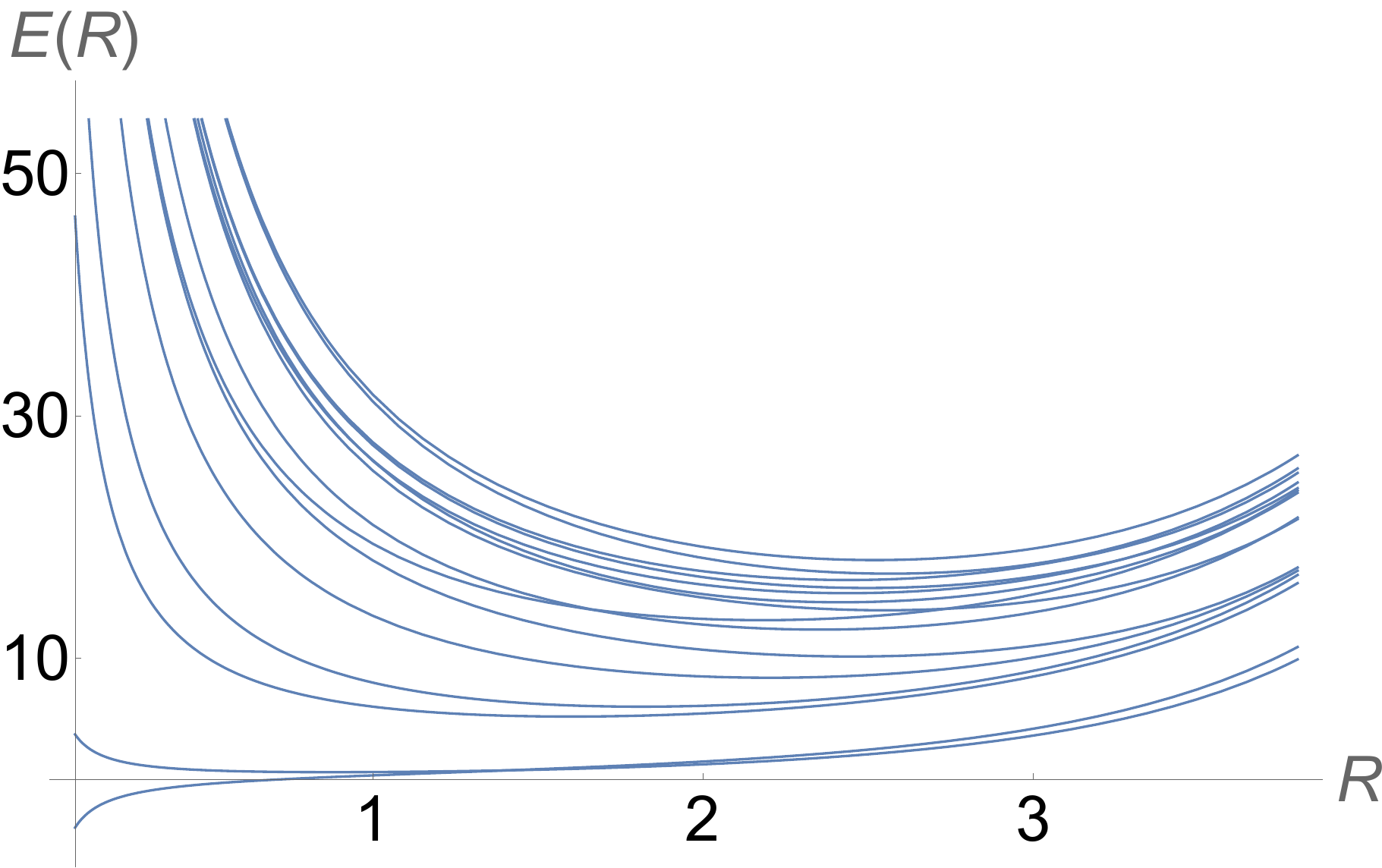}
    \subcaption{$\lambda_{1,3}=+0.1$}
    \label{positive13}
    \end{minipage} &
    \begin{minipage}[t]{0.6\hsize}
    \centering
    \includegraphics[width=0.8\textwidth]{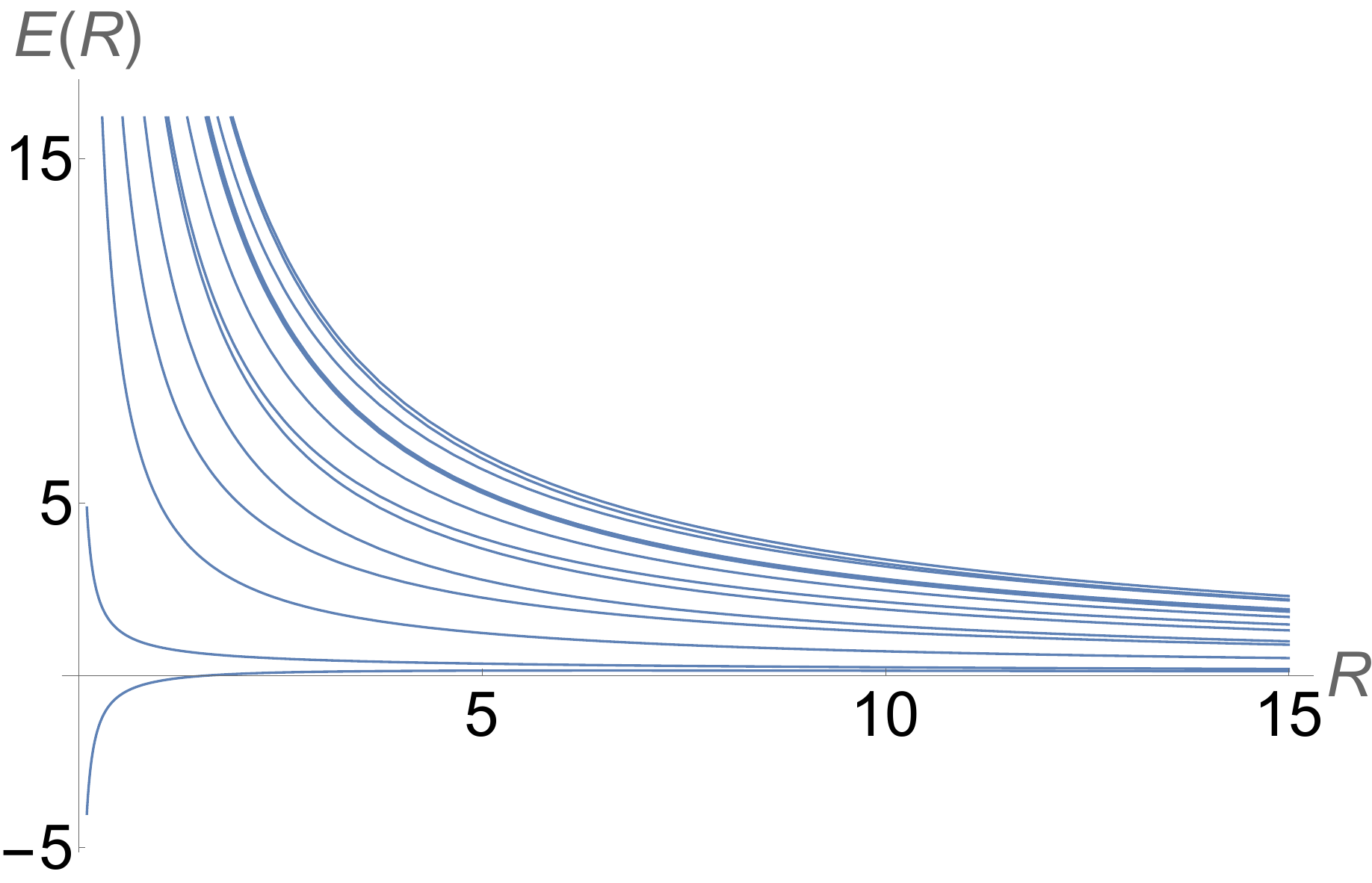}
    \subcaption{$\lambda_{1,3}=-0.1$}
    \label{negative13}
    \end{minipage} \\\hspace{-50pt}
    \begin{minipage}[t]{0.6\hsize}
    \centering
    \includegraphics[width=0.8\textwidth]{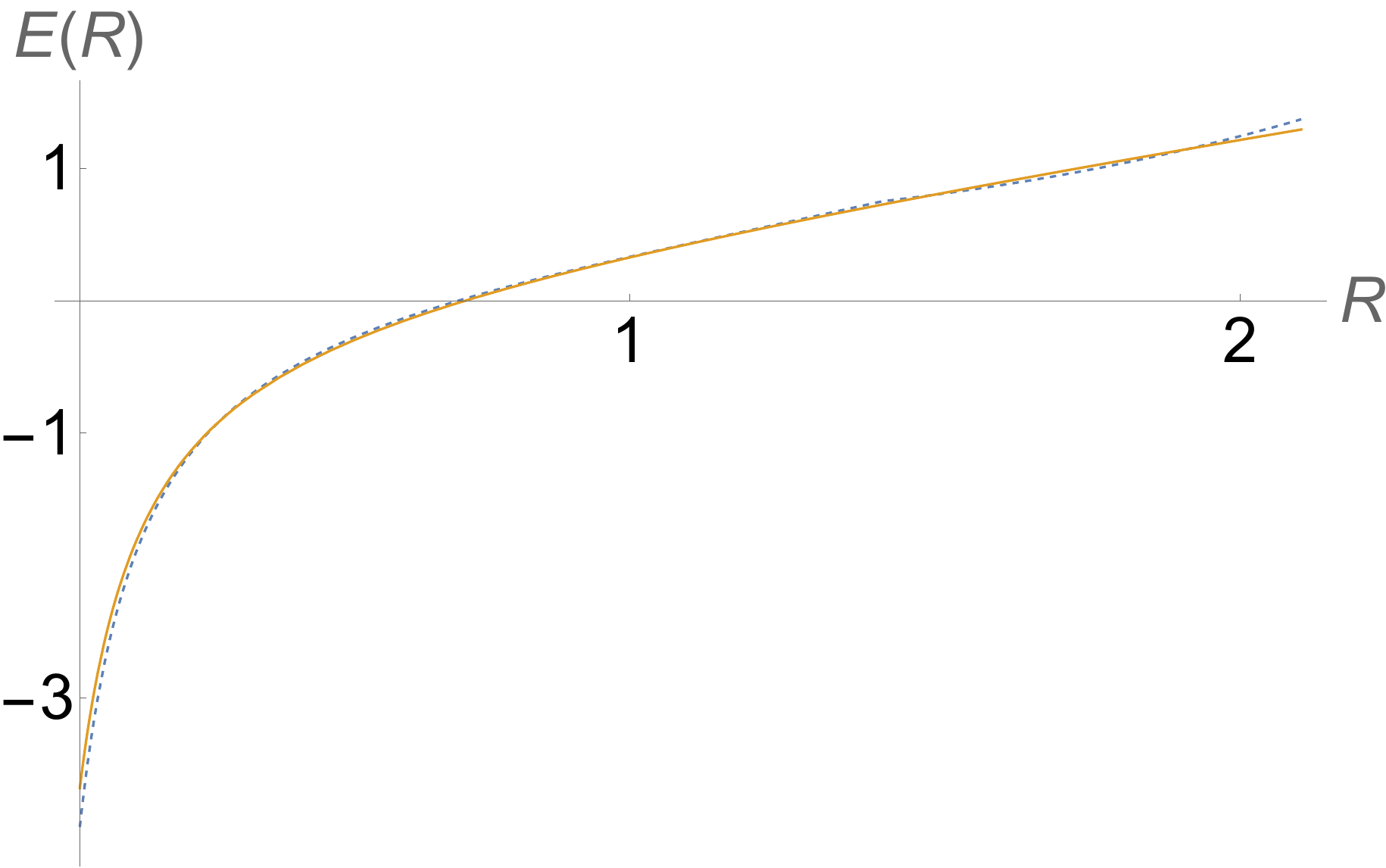}
    \subcaption{Computation of IR central charge}
    \label{fitting}
    \end{minipage} &
    \begin{minipage}[t]{0.6\hsize}
    \centering
    \includegraphics[width=0.8\textwidth]{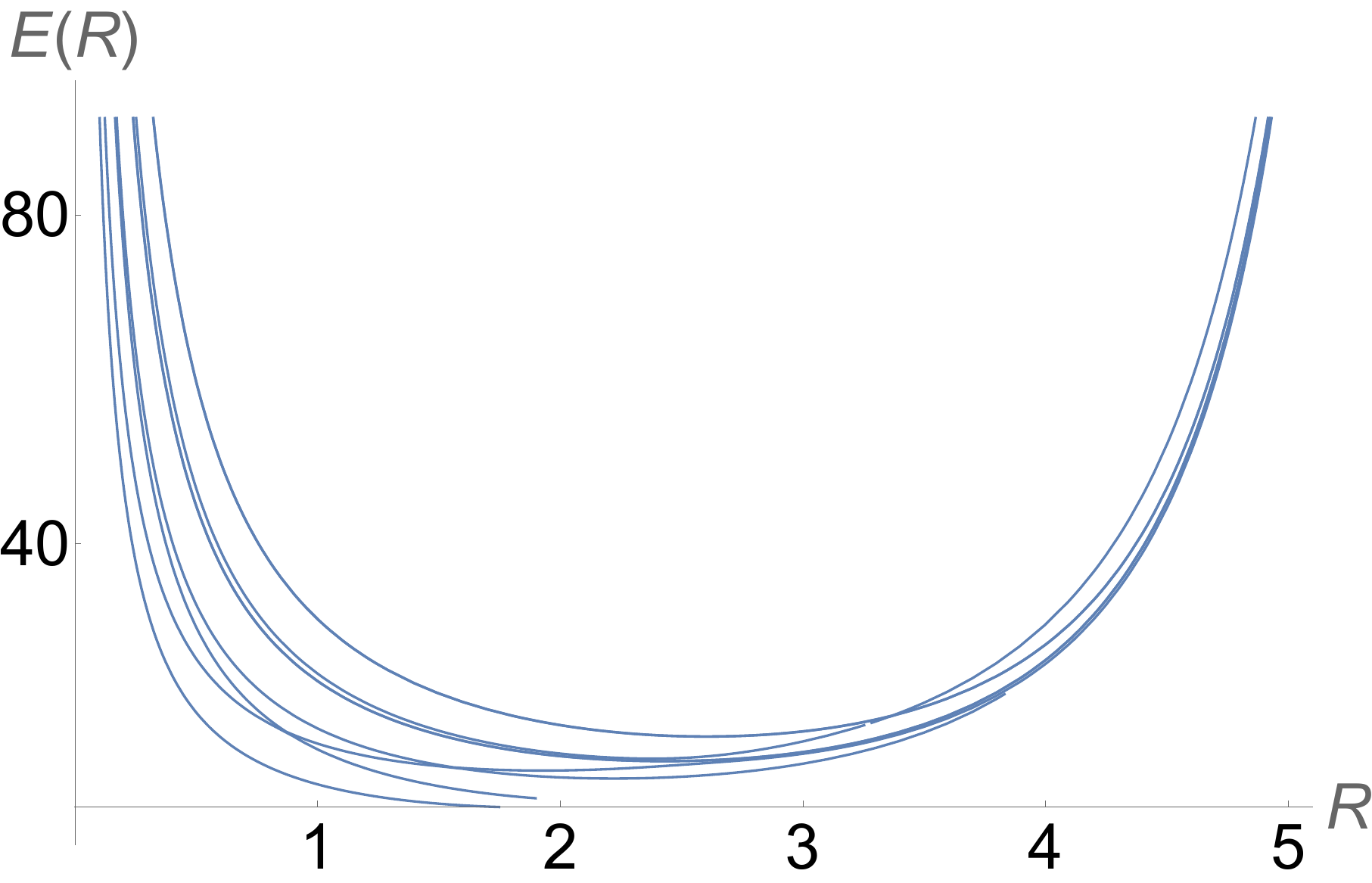}
    \subcaption{Spectra in the spin 3/2 sector}
    \label{spin3/2}
    \end{minipage} 
    \end{tabular}
    \caption{TCSA results: The spectra in the spin-zero sector for {\bf (a)} positive and {\bf (b)} negative relevant coupling $\lambda_{1,3}$. {\bf (c)} fits the ground state energy with ansatz $1/R+R$: the actual spectrum (dotted blue) and the fitting (solid orange). A fitting region is chosen to get the best fit. {\bf (d)} shows the spectra in the spin 3/2 sector for $\lambda_{1,3}=+0.1$.}\label{Fig:TCSA}
\end{figure}

The code needs two inputs: the basis of Verma modules and the operator product expansion (OPE) coefficients. The former is chiral in nature, and is the same as in the bosonic theories. The latter can be computed from $F$-symbols as in \cite{RW20}. We focus on the deformation by $\ep''\bar\ep''$ with the coupling constant $\lambda_{1,3}$. The TCSA results, $R$ vs. energy $E(R)$, are given in Figure \ref{Fig:TCSA}.
For $\lambda_{1,3}>0$, we see the two lowest spectra are separating from each other as we increase $R$, signaling a unique ground state. This strongly suggests the IR theory is conformal. As we argued in the previous section, the only candidate CFT is the fermionic $m=4$ minimal model. In fact, we can fit the ground state energy (IR central charge) following \cite{GW11}. The fitting result is given in figure \ref{fitting}.
The central charge from the fitting is
\begin{equation}
    c_\text{IR}^\text{numerical}=0.717383,\label{cIR}
\end{equation}
which is close to that of the fermionic $m=4$ minimal model $c=\frac{7}{10}$. The numerical result strongly supports our RG flow with emergent SUSY. We further apply the TCSA to the fermionic states.\footnote{The TCSA for the fermionic (half-integer spin) sector can be viewed as the TCSA for the defect Hilbert space of the ${\mathbb Z}_2$ line in the bosonic $m=5$ minimal model. We thank Shu-Heng Shao and Yifan Wang for a discussion on this point.} In particular, the spin $3/2$ states have spectra \ref{spin3/2}. By fitting curves, we find scaling dimensions
\begin{equation}
    \Delta=1.55239,\quad1.88536,\quad2.57076,\quad3.55231,\quad4.75231,\quad4.75231,\label{dims=3/2}
\end{equation}
which can respectively be identified with the operators $\ep''$, $L_{-1}\ep'\bar\ep$, $L_{-2}\ep\bar\ep'$, $L_{-2}\bar L_{-1}\ep'\bar\ep$, $L_{-3}\bar\ep''$, $L_{-3}\bar L_{-1}\ep\bar\ep'$ in the fermionic $m=4$ minimal model. The existence of half-integer spin states rules out the possibility of the IR theory being the bosonic $m=4$ minimal model.

Let us also look at the other sign of the coupling, $\lambda_{1,3}<0$. In this case, one can see a beautiful doubly-degenerated ground states, consistent with the surviving $W,(-1)^FW$ lines.\footnote{One may wonder whether the third lowest spectrum is a ground state or not. However, $\text{GSD}=3$ conflicts with the even-degeneracy imposed by the analytic constraints.} Our TCSA result suggests that $\lambda_{1,3}<0$ triggers an RG flow to the IR TQFT with $\text{GSD}=2$.

\section{Discussion}

We show that the fermionic $m=5$ minimal model deformed by the least relevant operator $\ep''\bar\ep''$ with a positive coupling constant flows to the fermionic $m=4$ minimal model. The IR emergent SUSY $\mcal N=1$ has ${\mathbb Z}_2\times {\mathbb Z}_2$ $R$-symmetry generated by the emergent q-type simple TDLs $R$ and $(-1)^FR$. Such a flow with emergent SUSY may open a door to discover SUSY experimentally.\footnote{See, say, \cite{GV12} for an attempt in this direction.}

The main tools we employ to solve the RG flow are constraints imposed by the TDLs, as well as a numerical method, TCSA. We showed their power by applying these to RG flows from non-Lagrangian theories. They can also be applied to study the RG flows from the fermionic $m=5$ minimal model induced by the relevant operators $\ep'\bar\ep'$ and $\ep\bar \ep$ (supplemental material \ref{theother2}), and the RG flow from the ferminonic $m=4$ model induced by the operator $\ep'\bar\ep'$ (supplemental material \ref{appendixC:m4tom3}). We believe the TDLs' analysis to constrain RG flows can be applied to general fermionic minimal models as well as other two-dimensional fermionic CFTs.

\section*{Acknowledgement}
We thank Jin Chen, Ling-Yan Hung, Shu-Heng Shao, Qing-Rui Wang for discussions.
We thank Shu-Heng Shao and Yifan Wang for comments on a preliminary draft.

\appendix
\setcounter{section}{0}
\renewcommand{\thesection}{\Alph{section}}
\setcounter{equation}{0}
\renewcommand{\theequation}{\Alph{section}.\arabic{equation}}

\section{TDLs in fermionic $m=3,4,5$ minimal models}\label{TDLsinm=345}
In this supplemental material, we explain how to find TDLs in fermionic CFTs via a (modified) Cardy condition \cite{C86,PZ00}.

The procedure works as follows. One first postulates the actions of a TDL $\mcal L$ on all the primary operators, and computes the twisted partition function
\begin{equation}
    Z^{\mcal L}(\tau,\bar\tau)=\tr\left( \widehat{\mcal L}q^{L_0-\frac{c}{24}}\bar q^{\widetilde L_0-\frac{c}{24}}\right)\,.
\end{equation}
Next, we take the modular $S$-transformation of the twisted partition function $Z^{\cal L}(-\frac{1}{\tau},-\frac{1}{\bar \tau})$. If the postulated TDL $\cal L$ is legitimate, then $Z^{\cal L}(-\frac{1}{\tau},-\frac{1}{\bar \tau})$ can be interpreted as a partition function of the defect Hilbert space $\mcal H_{\cal L}$, and should admit a $q$ and $\bar q$-expansion with natural number coefficients. Let us demonstrate this condition in RCFTs, where the number of primary operators is finite. Their characters $\chi_i$ combine to give the full partition function $Z=\sum_{i,j}N_{ij}\chi_i\bar\chi_j$. The coefficients $N_{ij}$'s should be natural numbers for the Hilbert space to have a physical interpretation. We place a theory on a torus. Suppose we insert a putative TDL $\mcal L$ along a time slice, and get the twisted partition function
$Z^{\mcal L}=\sum_{i,j}\left(M_{\mcal L}\right)_{ij}\chi_i\bar\chi_j$. The mass matrix $\left(M_{\mcal L}\right)_{ij}$ encodes the information how $\mcal L$ acts on primaries. In order to find actions of TDLs (or equivalently consistent mass matrix $M_{\mcal L}$), we perform the modular $S$-transformation. Then, the line is now inserted along the time direction. In other words, a time slice gives a defect Hilbert space $\mcal H_{\mcal L}$. We can still expand the trace over the space in terms of characters; $Z_{\mcal L}=\sum_{i,j}(SM_{\mcal L}S^\dagger)_{ij}\chi_i\bar\chi_j$. However, for the Hilbert space to have a physical interpretation, the coefficients should be natural numbers:
\begin{equation}
    (SM_{\mcal L}S^\dagger)_{ij}\stackrel!\in\mbb N.\label{mtypeCardy}
\end{equation}

In fermionic CFTs, there are two types of simple TDLs: m-type and q-type \cite{J88,ALW17}.\footnote{The latter were also called ``Majorana object'' in \cite{GWW10,GK15,U16}.} In the language of super fusion category, the m- and q-type simple TDLs are defined by the dimensions of the endomorphism spaces:
\begin{equation}\label{mqtype}
    \End(\mcal L_\text{m-type})={\mathbb C}\,,\quad\End(\mcal L_\text{q-type})={\mathbb C}^{1|1}\,.
\end{equation}
The m-type TDLs satisfy the ordinary Cardy condition \eqref{mtypeCardy}. For the q-type TDLs, \eqref{mqtype} implies that the defect Hilbert space ${\cal H}_{\mcal L_{\text{q-type}}}$ is doubly degenerate, and also indicates an extra one-dimensional Majorana fermion living on the worldline of the q-type simple TDL. A Majorana fermion contributes $\sqrt2$ to a torus partition function, and we arrive the modified Cardy condition
\begin{equation}
    (SM_{\mcal L}S^\dagger)_{ij}\stackrel!\in \sqrt2 \mbb N.\label{qtypeCardy}
\end{equation}
Equipped with the method, we find TDLs in fermionic minimal models with $m=3,4,5$ cases.\footnote{One could solve (modified) Cardy conditions via brute-force calculation. Alternatively, one could use the simple (but heuristic) method found in \cite{KK}.}

\subsection{$m=3$}

The $m=3$ minimal model has four primary operators
\begin{equation}\label{eqn:m=3primaries}
    1\,,\quad \psi\,,\quad\bar\psi\,,\quad \psi\bar\psi\,.
\end{equation}
There are three characters with the holomorphic conformal weights $h=0$, $\frac{1}{2}$, and $\frac{1}{16}$. Let us denote the action of a TDL $\mcal L$ on the primary operators by a mass matrix $M_{\mcal L}$:
\begin{equation}
    M_{\mcal L}=\begin{pmatrix}a&b&0\\c&d&0\\0&0&0\end{pmatrix}.\label{ML}
\end{equation}
where the rows and columns are ordered as $h=(0,\frac{1}{2},\frac{1}{16})$. Only four components are non-zero because we only have four primaries \eqref{eqn:m=3primaries} in the fermionic $m=3$ minimal model. Since there are four variables, $a,b,c,d$, there are at most four TDLs. Conjugating the mass matrix with the modular $S$-matrix, we obtain
\[ SM_{\mcal L}S^\dagger=\begin{pmatrix}A&A&B\\A&A&B\\C&C&D\end{pmatrix}, \]
where
\[ A:=\frac{a+b+c+d}4,\quad B:=\frac{a-b+c-d}{2\sqrt2},\quad C:=\frac{a+b-c-d}{2\sqrt2},\quad D:=\frac{a-b-c+d}2. \]
Let us first look for invertible lines. (In this case, one can set $a=1$.) We first try to find m-type lines by imposing the ordinary Cardy condition (\ref{mtypeCardy}):
\[ A,B,C,D\in\mbb N. \]
Adding or subtracting $B$ and $C$, one obtains
\[ \frac{a-d}{\sqrt2}\in\mbb N,\quad\frac{-b+c}{\sqrt2}\in\mbb N. \]
Obviously, the identity line $a=b=c=d=1$ satisfies the condition, which gives one (trivial) TDL. A nontrivial line is given by
\[ a=1=d,\quad b=-1=c. \]
This is nothing but the fermion parity $(-1)^F$.

Next, let us search for q-type lines. In this case, we impose the modified Cardy condition (\ref{qtypeCardy}). Adding or subtracting $B$ and $C$, we obtain
\[ \frac{a-d}{\sqrt2}\in\sqrt2\mbb N,\quad\frac{-b+c}{\sqrt2}\in\sqrt2\mbb N. \]
These conditions give two choices:
\[ a=1=b,\quad c=-1=d, \]
or
\[ a=1=c,\quad b=-1=d. \]
The former is nothing but $(-1)^{F_R}$ and the latter is $(-1)^{F_L}$. In total, we obtained four lines, and we can stop. The nontrivial lines have actions
\begin{table}[H]
\begin{center}
\begin{tabular}{c|c|c|c|c}
	&$1$&$\psi$&$\bar\psi$&$\psi\bar\psi$\\\hline
	$\widehat{(-1)^F}$&1&$-1$&$-1$&1\\
	$\widehat{(-1)^{F_L}}$&1&$-1$&1&$-1$\\
	$\widehat{(-1)^{F_R}}$&1&1&$-1$&$-1$
\end{tabular}.
\end{center}
\end{table}
One can also read off spin contents associated to each defect Hilbert space $\mcal H_{\mcal L}$ from the conjugated mass matrix $SM_{\mcal L}S^\dagger$:
\begin{align*}
    \mcal H_{(-1)^F}:\quad&s\in\{0\},\\
    \mcal H_{(-1)^{F_L}}:\quad&s\in\{\frac1{16},-\frac7{16}\},\\
    \mcal H_{(-1)^{F_R}}:\quad&s\in\{-\frac1{16},\frac7{16}\}.
\end{align*}

\subsection{$m=4$}
\label{sec:TDL_in_m=4}
The procedure can be repeated for $m=4$. Since the computations go through exactly the same (although it is a bit tedious), we only present the results. The theory has eight TDLs in total. The primitive lines (generators of the fusion ring) have actions
\begin{table}[H]
\begin{center}
\begin{tabular}{c|c|c|c|c|c|c|c|c}
	&$1$&$\ep\bar\ep$&$\ep'\bar\ep'$&$\ep''\bar\ep''$&$\ep''$&$\bar\ep''$&$\ep\bar\ep'$&$\ep'\bar\ep$\\\hline
	$\widehat{(-1)^F}$&1&1&1&1&$-1$&$-1$&$-1$&$-1$\\
	$\widehat W$&$\zeta$&$-\zeta^{-1}$&$-\zeta^{-1}$&$\zeta$&$\zeta$&$\zeta$&$-\zeta^{-1}$&$-\zeta^{-1}$\\
	$\widehat R$&1&$-1$&1&$-1$&$-1$&1&$-1$&1
\end{tabular}.
\end{center}
\end{table}
The spin contents associated to each defect Hilbert space are given by
\begin{align*}
    \mcal H_{(-1)^F}:\quad&s\in\{0\},\\
    \mcal H_W:\quad&s\in\{0,\pm\frac25,\pm\frac1{10}\},\\
    \mcal H_R:\quad&s\in\{-\frac1{16},\frac7{16}\}.
\end{align*}

\subsection{$m=5$}
\label{sec:TDL_in_m=5}
Finally, we repeat the same computation for the $m=5$ case.  By solving the ordinary Cardy condition, we find 10 TDLs in total. The action of the primitive lines are listed in Table \ref{TDLactionm=5}.
The spin contents associated to each defect Hilbert space are given by
\begin{align*}
    \mcal H_{(-1)^F}:\quad&s\in\{0\},\\
    \mcal H_W:\quad&s\in\{0,\pm\frac12,\pm\frac25,\pm\frac1{10}\},\\
    \mcal H_N:\quad&s\in\{\pm\frac18,\pm\frac38,\pm\frac1{24},\pm\frac{11}{24}\},\\
    \mcal H_M:\quad&s\in\{0,\pm\frac12,\pm\frac13\}.
\end{align*}

\section{The other two RG flows}\label{theother2}
For completeness, in this supplemental material, we consider the other two RG flows triggered by the relevant operators $\phi_{21,21}=\ep'\bar\ep'$ and $\phi_{33,33}=\ep\bar\ep$. 

\begin{figure}[H]
\begin{minipage}[b]{0.45\linewidth}
    \centering
    \includegraphics[keepaspectratio, scale=0.28]{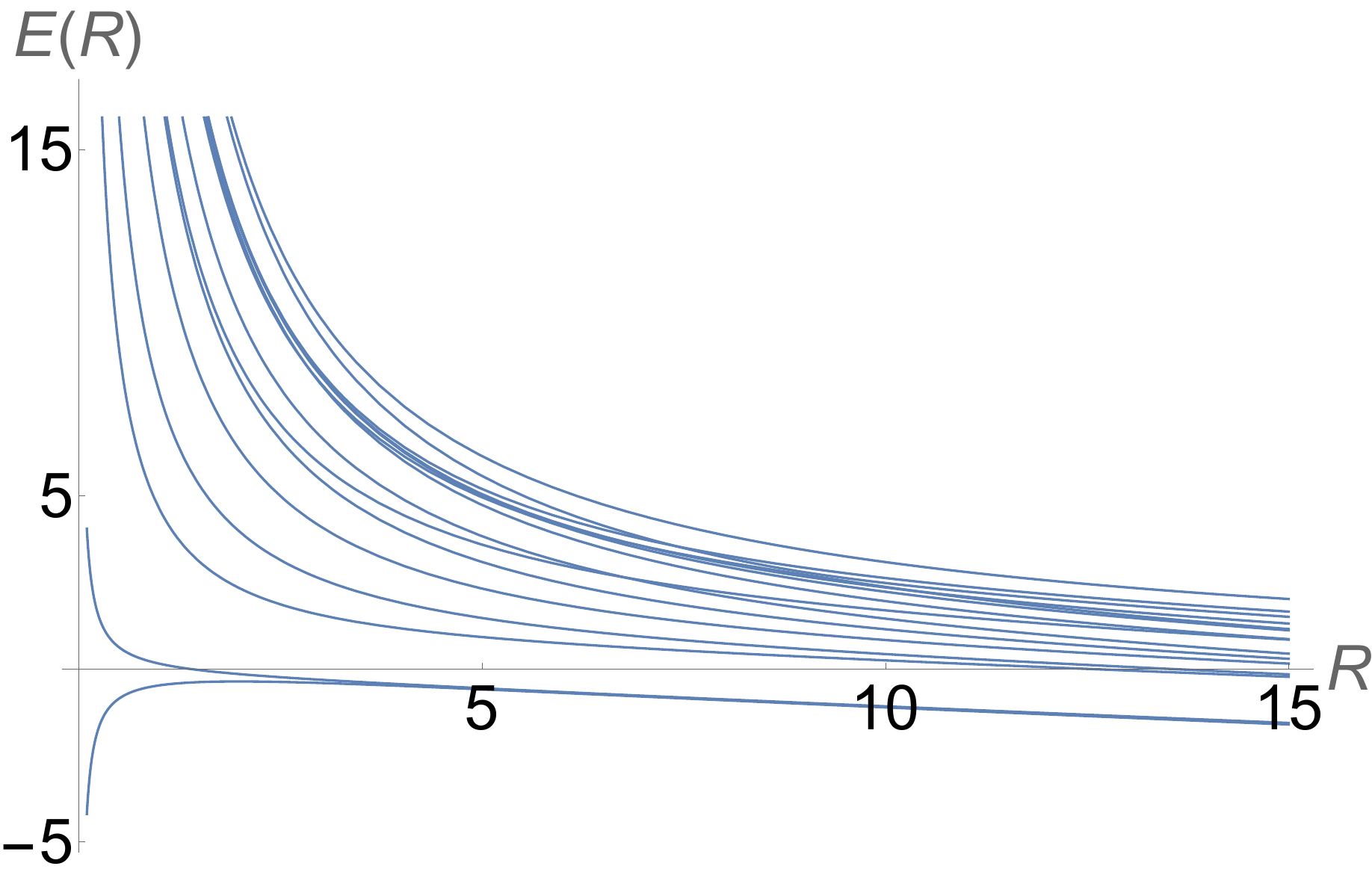}
    \subcaption{$\lambda_{2,1}=+0.1$}
  \end{minipage}
  \hspace{50pt}\begin{minipage}[b]{0.45\linewidth}
    \centering
    \includegraphics[keepaspectratio, scale=0.28]{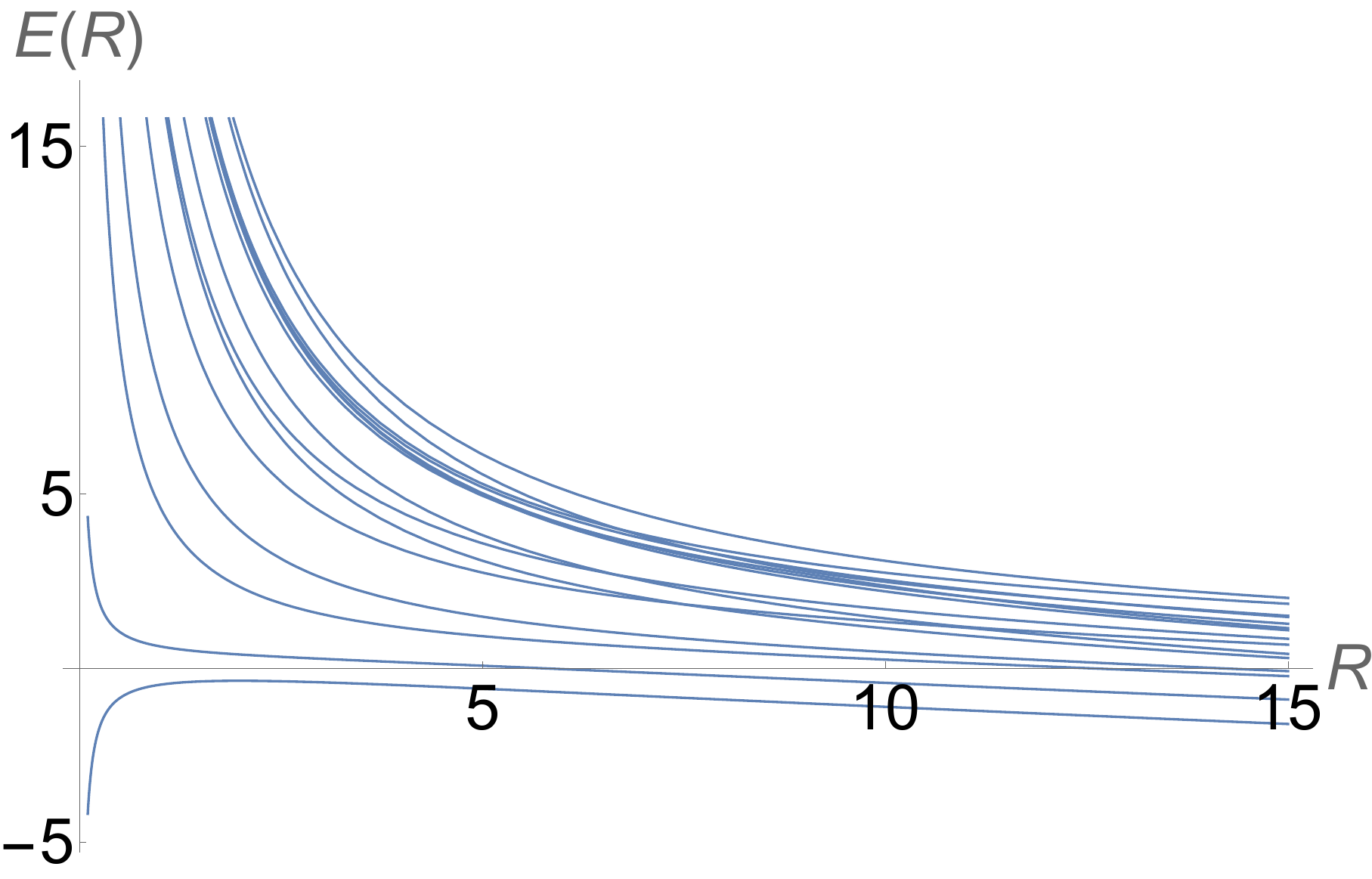}
    \subcaption{$\lambda_{2,1}=-0.1$}
  \end{minipage}
  \caption{TCSA results: The spectra in the spin-zero sector for the {\bf (a)} positive and {\bf (b)} negative relevant coupling $\lambda_{2,1}$.}\label{Fig:TCSA21}
\end{figure}

We start with the RG flow triggered by $\ep'\bar\ep'$, which commutes with the primitive lines $\{I,(-1)^F,M\}$. One finds no lower CFT can satisfy the constraints --- the matching of the quantum dimensions, fusion ring \eqref{eqn:fusionrulem=5}, $F$-moves, and (topological) $S$-matrix --- of the three TDLs. Hence, the IR theory cannot be a CFT, and is necessarily gapped. It can be trivial because all quantum dimensions are natural numbers. To fix the ground state degeneracy (GSD), we employ the TCSA. The numerical results are given in Figure \ref{Fig:TCSA21}, which suggest
\begin{equation}
    \text{IR theory}=\begin{cases}\text{TQFT with GSD}=2&(\lambda_{2,1}>0),\\
    \text{TQFT with GSD}=1&(\lambda_{2,1}<0).\end{cases}\label{phi21IRphases}
\end{equation}

Next, let us study the RG flows preserving only the TDLs $\{I,(-1)^F\}$, which can in general be triggered by any linear combination $\lambda_{3,3}\ep\bar\ep+\lambda_{2,1}\ep'\bar\ep'+\lambda_{1,3}\ep''\bar\ep''$. Since the number of surviving TDLs are small, the analytic constraints we can draw is weak. In fact, we cannot rule out any scenario, and the IR theory can be gapped (possibly degenerate), or a CFT. For the CFT scenario, from the $c$-theorem, the bosonic and fermionic minimal models with $m=3,4$ are allowed. However, the spin constraint \eqref{spinmatch} rules out bosonic options just as the flow discussed in the main text. Hence, the possible IR CFTs are the fermionic minimal models with $m=3,4$. In these scenarios, the $(-1)^F$ line flows to the $(-1)^F$ lines in the IR. In general, we expect that the CFTs or the TQFTs with degenerate vacua only show up when we fine tune the coupling constants $\lambda_{3,3}$, $\lambda_{2,1}$, $\lambda_{1,3}$; otherwise, the IR phase is trivially gapped.

\begin{figure}[H]
\begin{minipage}[b]{0.45\linewidth}
    \centering
    \includegraphics[keepaspectratio, scale=0.28]{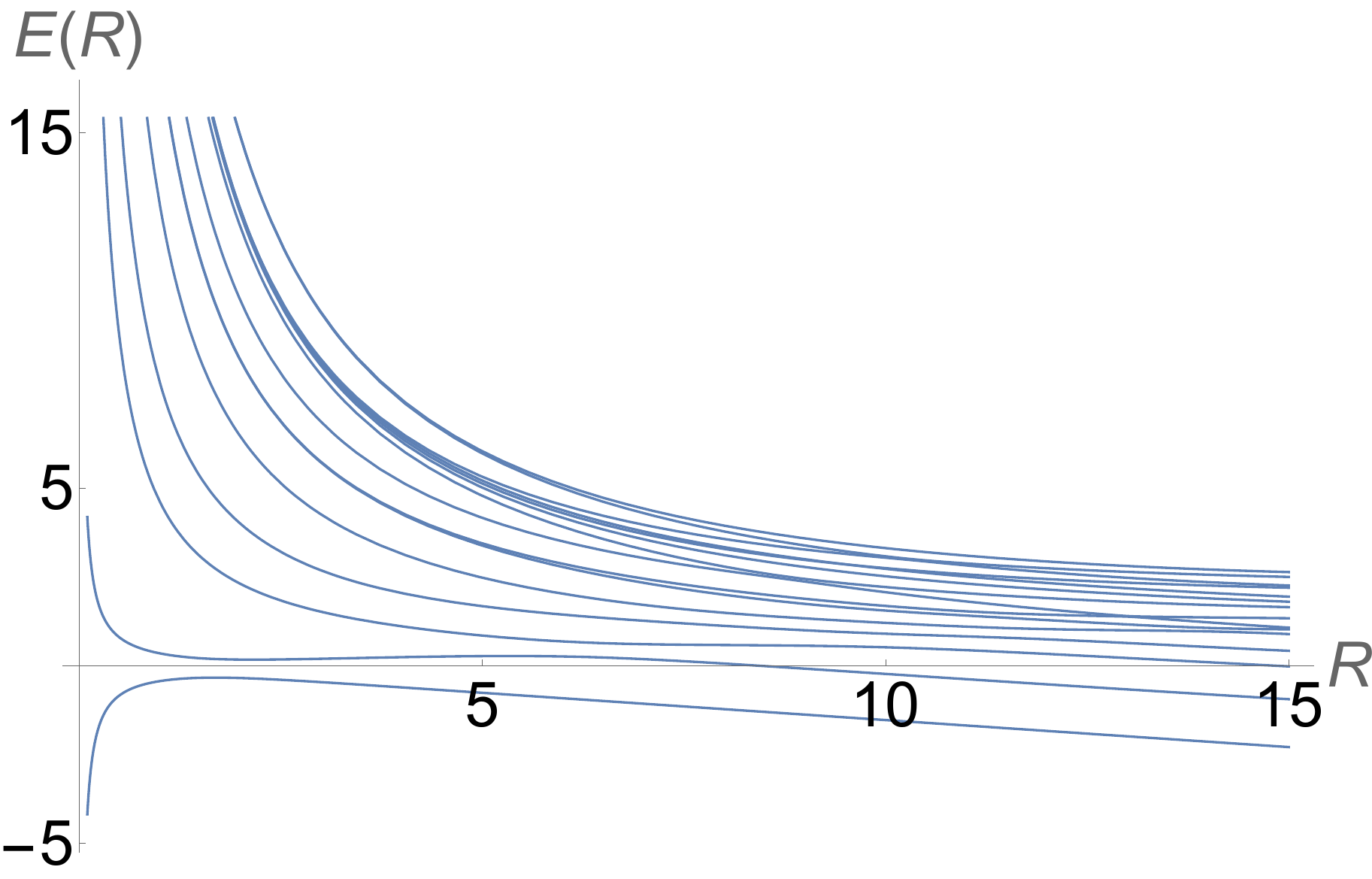}
    \subcaption{$\lambda_{3,3}=+0.1$}
  \end{minipage}
  \hspace{50pt}\begin{minipage}[b]{0.45\linewidth}
    \centering
    \includegraphics[keepaspectratio, scale=0.28]{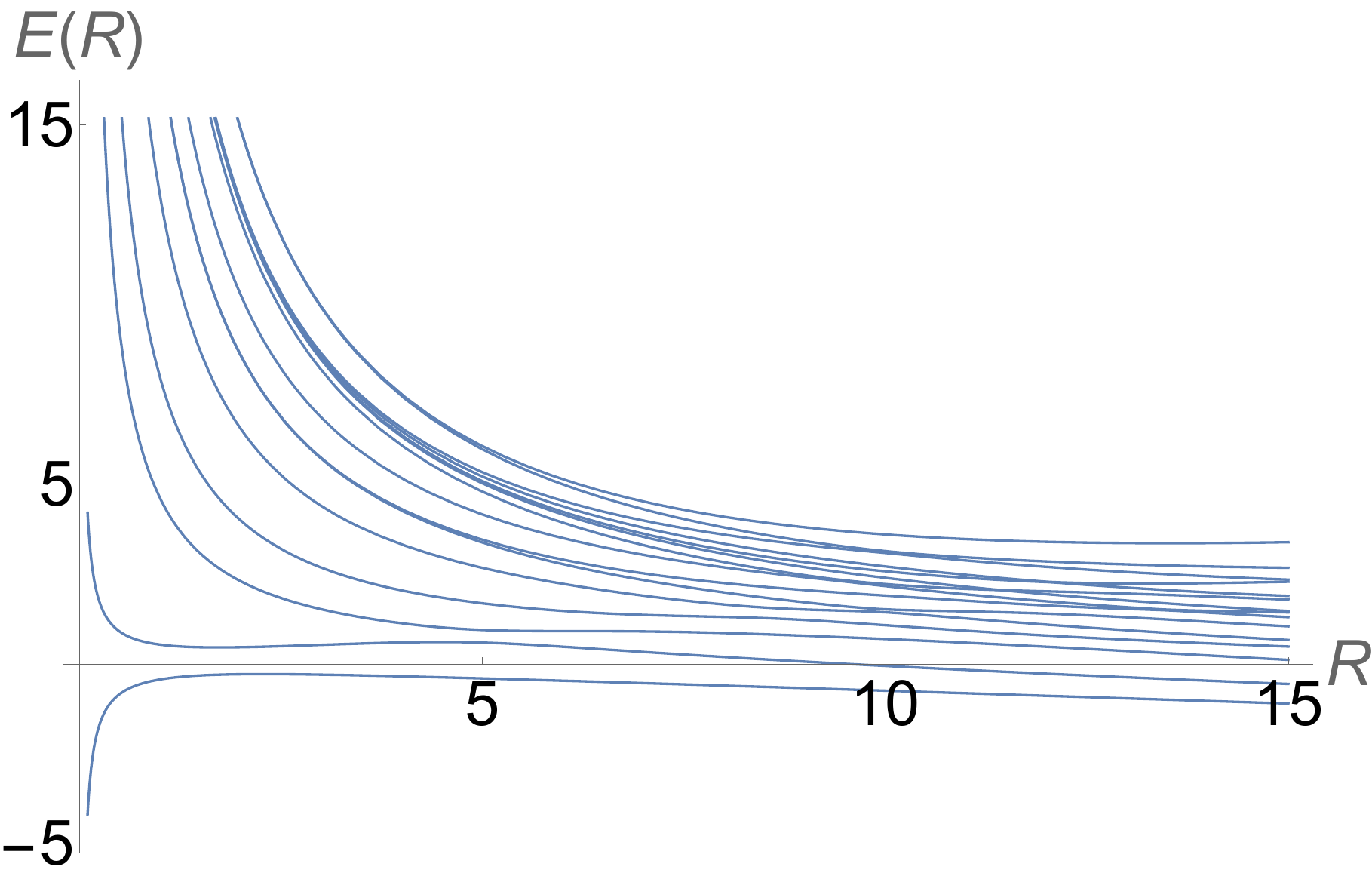}
    \subcaption{$\lambda_{3,3}=-0.1$}
  \end{minipage}
  \caption{TCSA results: The spectra in the spin-zero sector for the {\bf (a)} positive and {\bf (b)} negative relevant coupling $\lambda_{3,3}$.}\label{Fig:TCSA33}
\end{figure}

For concreteness, let us consider the case $\lambda_{2,1}=0=\lambda_{1,3}$. The results are given in Figure \ref{Fig:TCSA33}, which suggest $\text{GSD}=1$ for both signs of the relevant coupling $\lambda_{3,3}$. The behavior is closer to gapped phases than to conformal phases. In order to check the phase, we also fit the ground state energy. However, we found the numerical results are not close to neither $\frac7{10}$ nor $\frac12$. Therefore, we believe that the IR theories are trivially gapped.

\section{RG flow from fermionic $m=4$ to $m=3$}
\label{appendixC:m4tom3}
\begin{figure}[H]
\begin{minipage}[b]{0.45\linewidth}
    \centering
    \includegraphics[keepaspectratio, scale=0.28]{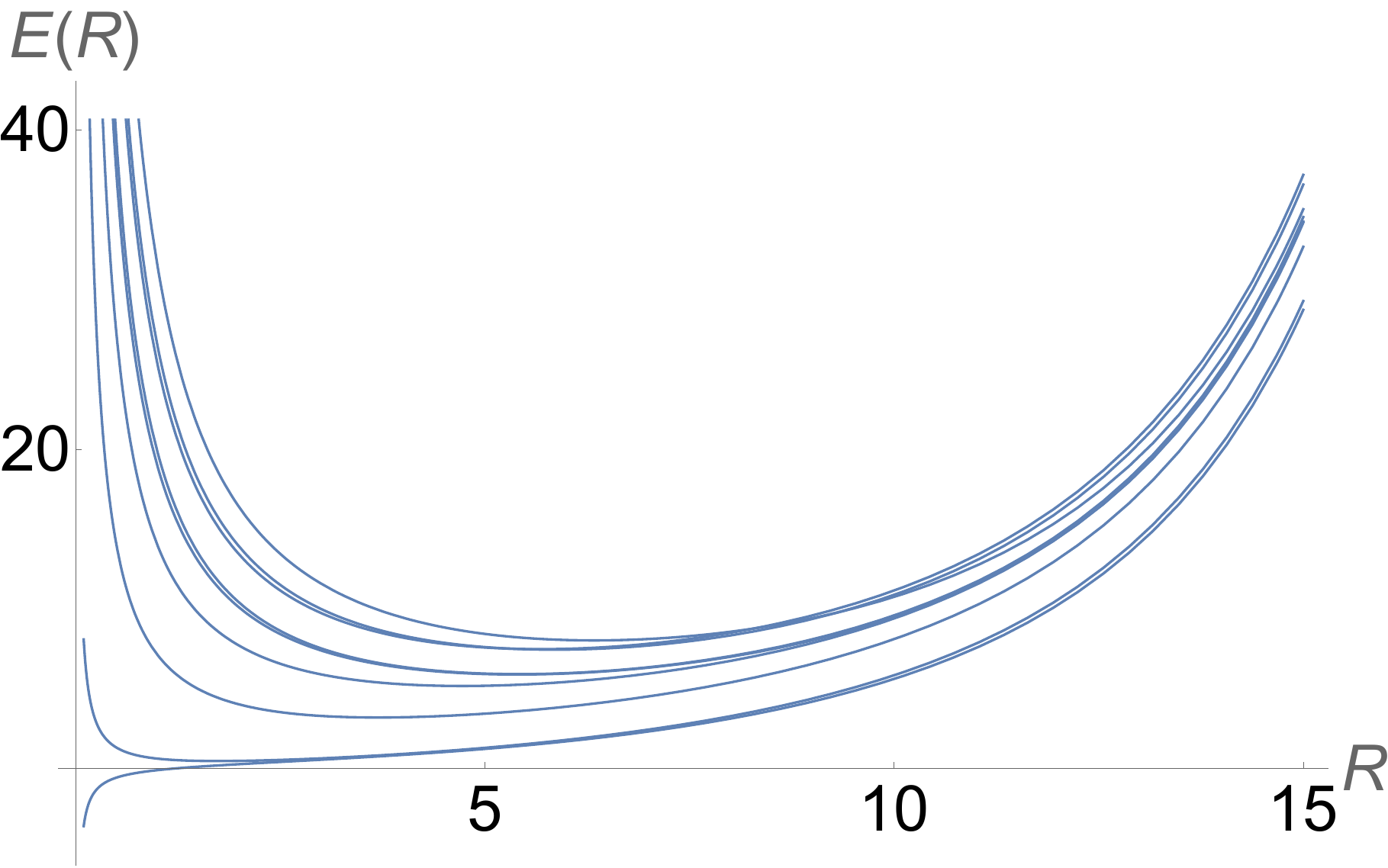}
    \subcaption{$\lambda_{1,3}=+0.1$}
  \end{minipage}
  \hspace{50pt}\begin{minipage}[b]{0.45\linewidth}
    \centering
    \includegraphics[keepaspectratio, scale=0.28]{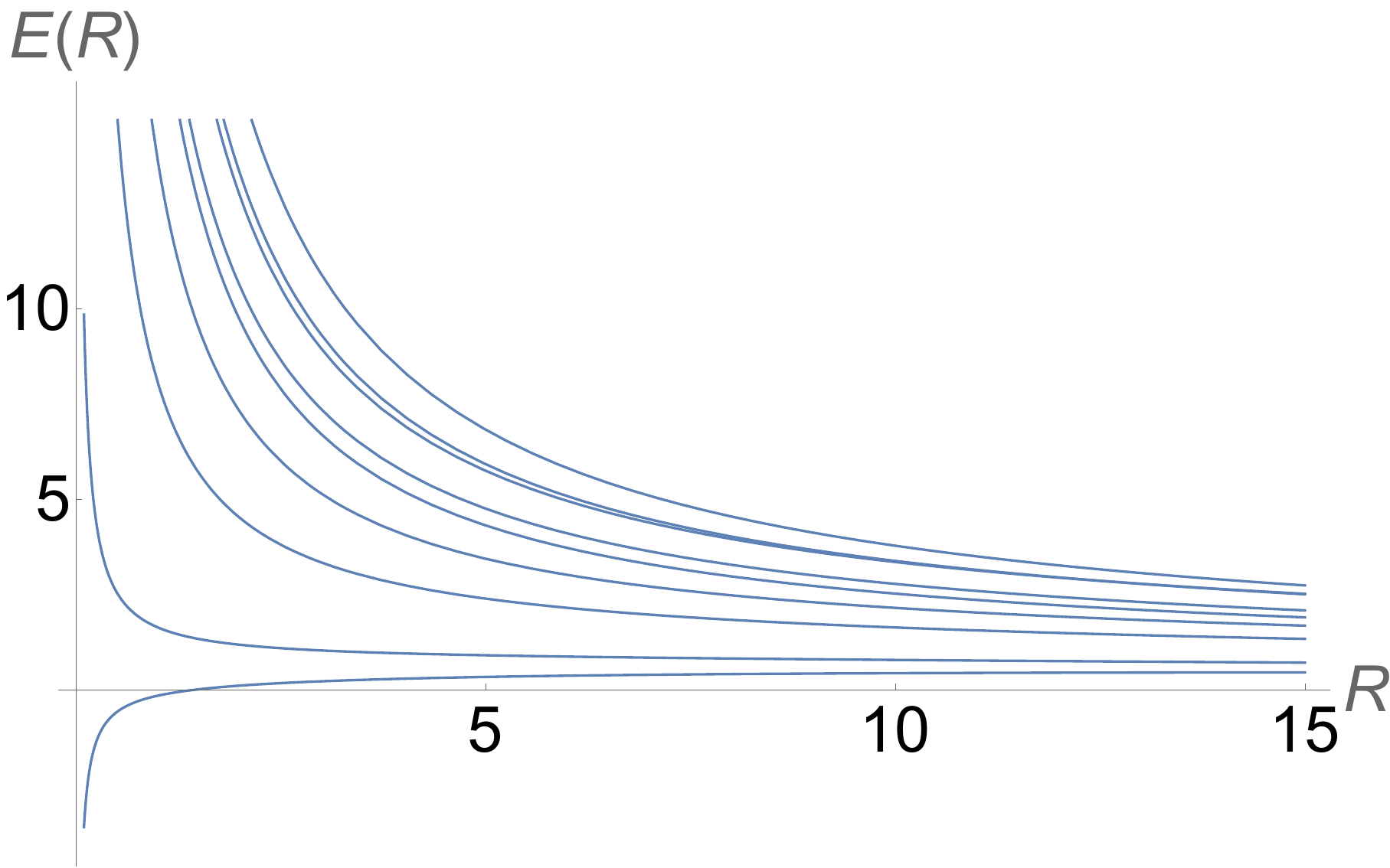}
    \subcaption{$\lambda_{1,3}=-0.1$}
  \end{minipage}
  \caption{TCSA results: The spectra in the spin-zero sector for the {\bf (a)} positive and {\bf (b)} negative relevant coupling $\lambda_{1,3}$ in the deformed fermionic $m=4$ minimal model.}\label{fig:TCSA13m=4}
\end{figure}

In this supplemental material, we consider the RG flow from the fermionic $m=4$ minimal model to the fermionic $m=3$ minimal model. We argue that such an RG flow can be triggered by the relevant operator $\phi_{13,13}=\ep'\bar\ep'$, which commutes with four TDLs $\{I,(-1)^F,R,(-1)^FR\}$. All the lines have integer quantum dimensions, but the lines $R$ and $(-1)^F R$ have non-trivial $F$-symbols. Hence, the trivial phase is ruled out. The IR theory can thus be a TQFT with $\text{GSD}>1$, or a CFT. Let us consider the CFT scenario first. The $c$-theorem tells us that the only candidates are bosonic or fermionic $m=3$ minimal models. However, the former is ruled out because it does not have three (nontrivial) invertible TDLs.\footnote{The spin constraint \eqref{spinmatch} also kills the possibility.} Therefore, the only candidate is the fermionic $m=3$ minimal model. Next, let us also look at the TQFT scenario. In this case, the surviving lines are not strong enough to constrain GSD. Hence, we resort to the TCSA. The results given in Figure \ref{fig:TCSA13m=4} suggest
\begin{equation}
    \text{IR theory}=\begin{cases}\text{fermionic $m=3$ minimal model}&(\lambda_{1,3}>0),\\
    \text{TQFT with GSD}=2&(\lambda_{1,3}<0).\end{cases}\label{m=4+phi13}
\end{equation}

\end{document}